# FPGA Meets Breadboard: Integrating a Virtual Breadboard with Real FPGA Boards for Remote Access in Digital Design Courses

Shuowei Li[1[0000-0003-4934-4691]] , Heran Wang[1[0000-0003-1207-9011]], Luis Rodriguez-Gil[2[0000-0003-3611-1418]] Pablo Orduña[2[0000-0003-4024-3624]] Rania Hussein[1[0000-0002-2859-9401]]

[1] University of Washington, Seattle WA 98195, USA

{lis26, hw99, rhussein}@uw.edu

[2] LabsLand, Bilbao, Spain

{luis, pablo}@labsland.com

**Abstract.** With the transition to remote instruction, new modalities for conducting hands-on labs are needed. In particular, courses that entail major hardware components faced challenges in making the hardware available for students in a reliable and sustainable way. This paper presents our experience in implementing a virtual breadboard feature to interface with real Field Programmable Gate Arrays (FPGAs) boards located on the University of Washington's campus, where students can access the boards remotely to complete their lab assignments.

## 1 Introduction

Traditionally, universities teach Field Programmable Gate Arrays (FPGAs) courses using on-campus labs, which have lab benches, PC/workstation, and other supporting equipment. In order to complete assignments, students need to be physically present in the labs to interact with these pieces of equipment. However, this approach is not feasible during the COVID-19 pandemic. A new teaching strategy based on a global network of remote laboratories is appealing in this era.

Remote laboratory facilities have never been a new topic. It has been used in teaching physics, biology, and other subjects [1][2][3]. However, the majority of them are stand-alone systems that cannot be shared with different facilities. In the field of engineering education, VISIR designs a shareable system to provide seamless access for students from various universities [4]. It allows students to explore electronics, sound, and vibration or propagate radio signals through an online platform. Even though the previous works demonstrate how applicable the educational remote laboratory facilities are, the state-of-the-art applications lack the ability to teach the student how to use General Purpose Input Output (GPIO) ports to communicate with the FPGA boards.

In particular, FPGA has been a target of remote laboratories in the literature for many years, in different initiatives in different universities [9][10][11][12]. In [8], a distributed FPGA laboratory with multiple instances in multiple locations is described.

*The first author and the second author contributed equally to this work

In this work, we added a new concept on top of [8]. This new concept intends to increase the scope of the lessons by providing an easy-to-use solution for students to work with the GPIO ports of the FPGA.

This paper introduces a feature that implements a virtual breadboard where students can use GPIO ports to interface with real FPGA boards that are accessed remotely through a global network called LabsLand [5][8]. This remote FPGA laboratory has been used to teach a junior-level course in digital design at the University of Washington during the autumn quarter of 2020. It presents a pilot project that intends to serve the needs of the course post-COVID-19. This paper presents our experience in transforming a lab that traditionally used physical breadboards and FPGA boards to an online modality. We demonstrate how the lab used to be conducted pre-Covid-19 and how it has been restructured to fit online instruction.

## 2 Background

The laboratory experiment highlighted in this paper is one out of 5 other labs that students are expected to complete for the course. This lab experiment is a refresher of the finite state machines and introduces students to the onboard general-purpose input/output (GPIO). The students achieve this goal during the process of designing a parking lot occupancy counter, which consists of two pairs of photosensors monitoring the activity of cars. For educational purposes, students use switches to mimic the output signal from the sensors. The signals are also displayed on two LEDs to make the signals visible while introducing more circuit components to the laboratory. This gives students opportunities to practice implementing breadboard circuits and interfacing GPIOs, which are essential skills for hardware design.

Traditionally, in the on-campus laboratory environment, each student receives a breadboard assembly in conjunction with a DE1-SoC FPGA, which is shown in Fig. 1. Students will implement three switches and two LEDs on the breadboard for this laboratory experiment. An example of breadboard circuits is shown in Fig. 2.

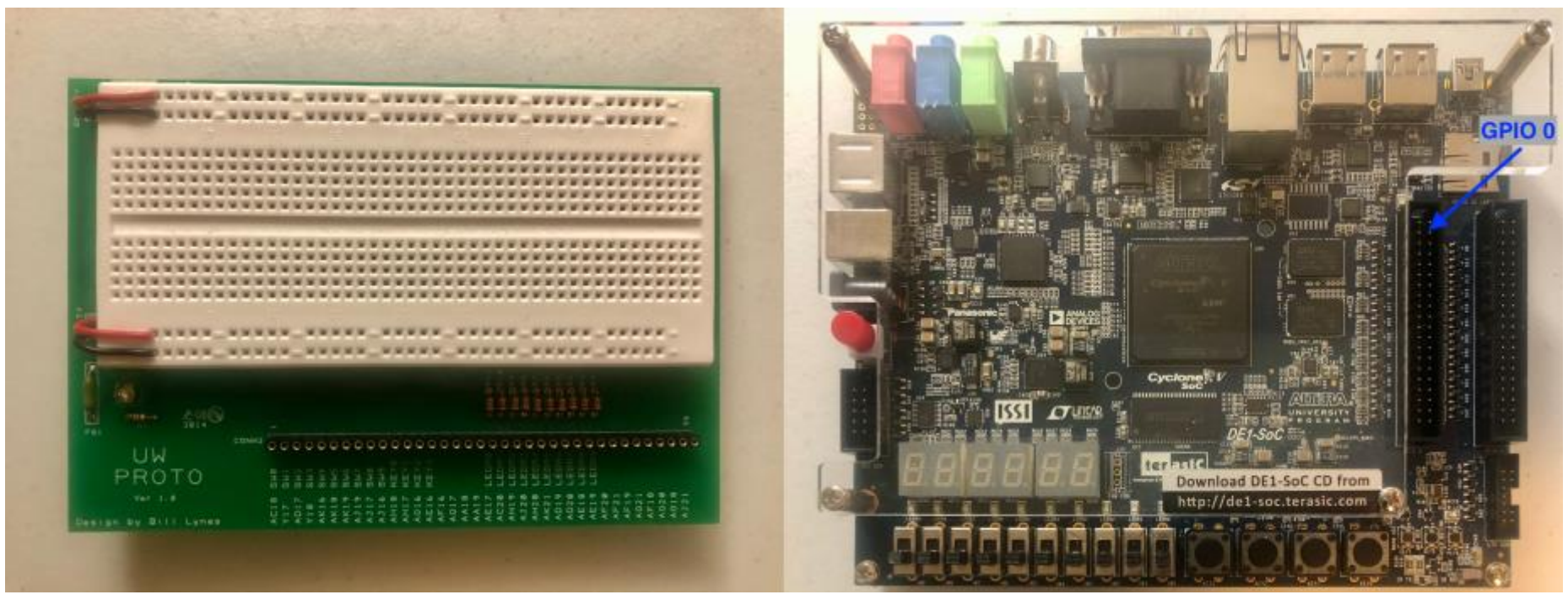

**Fig. 1.** The customized breadboard module (left) and the DE1-SoC FPGA (right). The breadboard module is connected to the FPGA board via the GPIO 0 header.

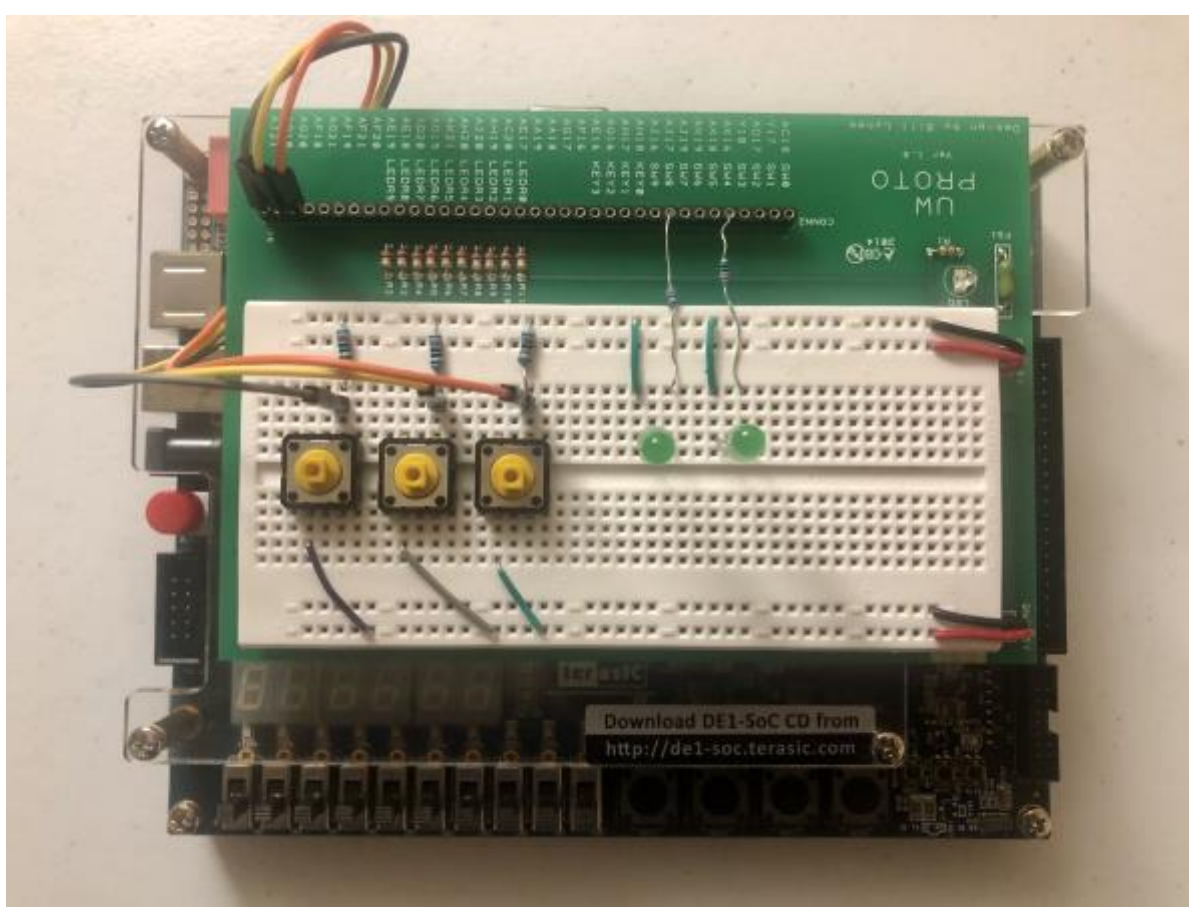

**Fig. 2.** The breadboard circuits that are required by the laboratory experiment.

However, the manufacture of these breadboard modules for large classes is time-consuming, and, most importantly, the physical laboratory kit is not accessible in the remote learning environment. Therefore, for this particular laboratory experiment, a remote laboratory solution that does not rely on the distribution of hardware is needed.

## 3 Technical Solution

### 3.1 User Perspective

To provide an uncompromising learning experience to students, we added a virtual breadboard interface to the remote laboratory facility LabsLand to simulate hands-on circuit building, providing students the opportunity to interface GPIOs in their SystemVerilog designs.

**Design Decisions for the Breadboard User Interface.** A breadboard configurator is added to the user interface switch to simulate the hands-on breadboarding experience. Students have access to several circuit components in the configurator, including three switches and two LEDs placed onto the virtual breadboard. We have a GPIO header that is internally connected to the DE1-SoC's JP1 in the backend of the system. Next to the breadboard's implementation area, we provide a pinout diagram for DE1-SoC's GPIO 0 (JP1) to help students understand the relationship between pin names and pin numbers and build the circuit. Fig. 3 depicts the default state of the virtual breadboard.

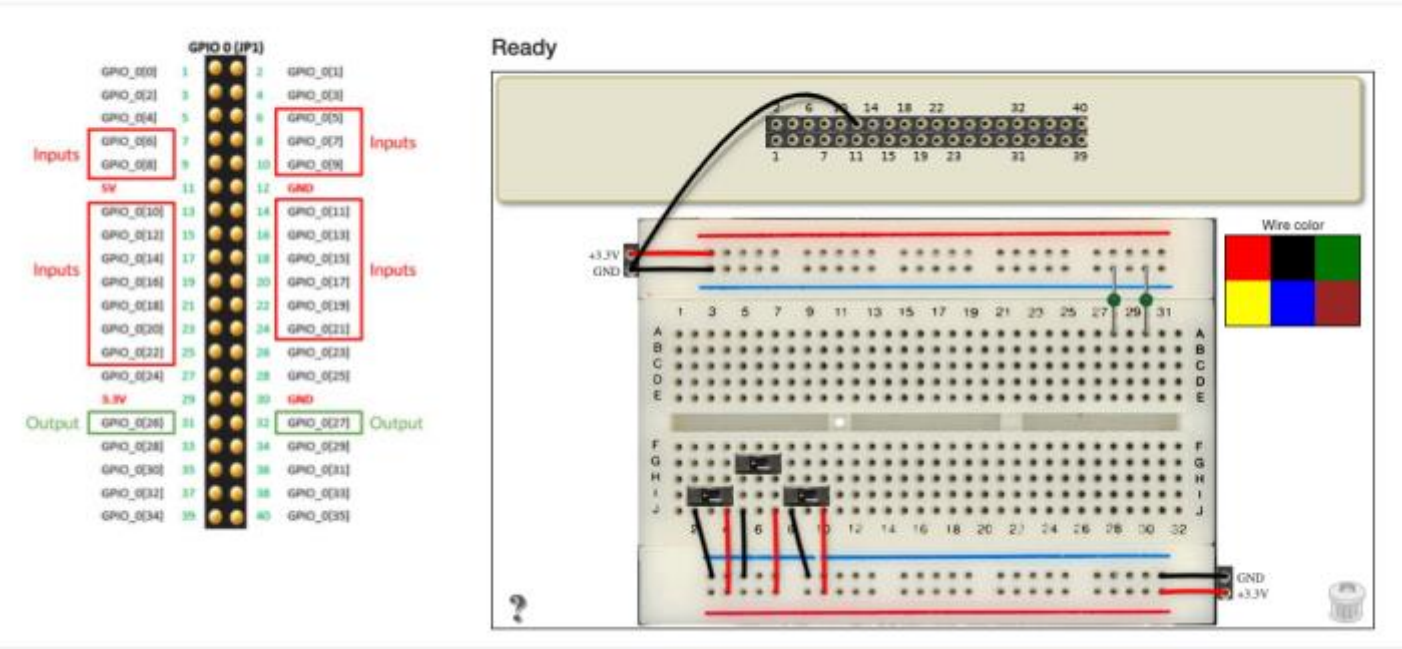

**Fig. 3.** The breadboard configurator on the LabsLand IDE.

Students will connect each circuit component to an appropriate GPIO pin of their choice for this particular laboratory experiment. To start with, students need to figure out which pin to use by referring to the pinout diagram to the left. As the pin names (e.g., *GPIO_0[20]*) are deliberately not shown on the actual GPIO header, students are forced to use the relationship between pin names and pin numbers, which ensures that all the essential steps for working with the actual boards are taken on the web interface as well. Then, students click on a wire color in the wire selector to choose a wire. They can use their mouse as a pen for drawing wires, connecting the GPIO header and the breadboard. All the connections on the breadboard will be automatically saved to emulate the real-life experience.

Students can then proceed to the remote laboratory facility LabsLand code editor interface to continue implementing the constructed virtual breadboard circuit. Once they synthesize and upload the project to FPGA, the user interface shown in

Fig. **4** will be presented to the students with all the previous changes made to the breadboard preserved. The students can toggle the switches and observe the LEDs' states along with other outputs on the DE1-SoC board, such as the HEX display, to verify their design. A tutorial specifically for the breadboard user interface is provided to the students to help them get familiar with the workflow.

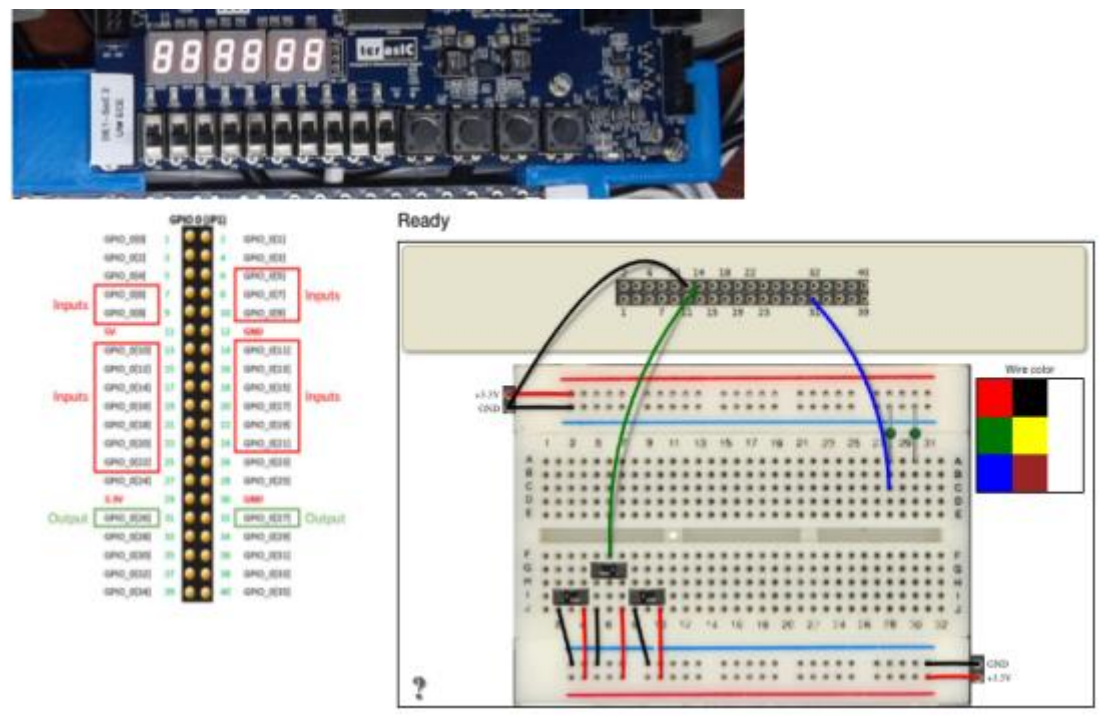

**Fig. 4.** The breadboard user interface for testing FPGA programs.

This workflow provides several benefits over other possible implementations. Firstly, it closely emulates the process of building a breadboard circuit in a physical laboratory setting, as it allows students to build the circuit before or in parallel with writing code. Secondly, by always saving the state of the breadboard for each student, it simulates the irreversible and nonvolatile nature of working with physical hardware. Thirdly, while providing unlimited time for building breadboard circuits, it also reduces the test time each student spends on the limited number of the FPGA hardware, reducing queueing time and providing a better learning experience for more students.

### 3.2 Technical Perspective

For the breadboard user interface, the Open Source VISIR[1] client has been used as an external library, particularly the breadboard component. On top of the breadboard component, a JavaScript code has been added that checks if certain points of the user interface are connected to the GPIO's and the switches and LED's on the breadboard. The system automatically detects faulty connections to help students connect what they need to connect to the particular GPIO.

The wiring phase is typically done in the code editor [8]. Once the breadboard is wired, whenever students use the FPGA, they have the same circuit built before, and they can still modify it once they are connected to the FPGA itself. Internally, each change of the wires or each change of the switches state is translated to a call to the server to send specific signals to the GPIO's. Once we detect any wire changing calls, the web browser will turn on and off the LEDs accordingly.

## 4 The FPGA with Breadboard Laboratory Experience at the UW

### 4.1 Contextualization

Due to the logistical difficulty of shipping physical DE1-SoC lab kits to students during the COVID-19 pandemic, all students in the Design Of Digital Circuits class at the UW in Autumn 2020 were asked to use the remote laboratory facility LabsLand along with the newly developed breadboard user interface. The remote laboratory does not require the shipping of any laboratory equipment and can be accessed anywhere and anytime. Before taking this class, all students are required to take a prerequisite junior-level Design of Digital Circuits and System course. Since the remote laboratory facilities just deployed for this offering. Students only have prior knowledge of using the physical DE1-SoC development board and breadboards.

---

[1] https://github.com/JohanZackrisson/visir_html5

### 4.2 Survey and Results

The students in Design Of Digital Circuits And Systems class were asked to complete an anonymous online survey to indicate their degree of agreement on certain statements based on a 5-point Likert scale [6], with one being “strongly disagree” and five beings “strongly agree.” Out of 60 students enrolled in the course, 55 students completed the survey. The survey questions were designed to evaluate their experience using the remote platform with the newly developed breadboard user interface. The survey and the corresponding quantitative data for statistical analysis are given in Table 1.

**Table 1.** The online survey, questions, and quantitative data (N = 55).

| # | Question | Min | Max | Mean | Std. Dev. | Var. |
|---|---|---|---|---|---|---|
| 1 | The pinout diagram shown on the Breadboard UI was helpful | 2 | 5 | 4.02 | 0.89 | 0.8 |
| 2 | Connecting wires on the breadboard UI was easy | 1 | 5 | 3.35 | 1.25 | 1.56 |
| 3 | Testing the breadboard circuit with my code was easy | 1 | 5 | 3.55 | 1.1 | 1.22 |
| 4 | The breadboard interface does not need improvement | 1 | 5 | 2.95 | 1.18 | 1.39 |
| 5 | The breadboard interface closely emulated the real physical breadboard | 1 | 5 | 3.67 | 1.12 | 1.26 |
| 6 | The Breadboard UI helped me understand how to use FPGA’s GPIOs | 1 | 5 | 3.64 | 1.08 | 1.16 |
| 7 | I was less afraid of damaging the FPGA board than when I work with circuits in the traditional lab | 1 | 5 | 3.58 | 1.46 | 2.14 |
| 8 | The Breadboard UI worked out without any problems | 1 | 5 | 3.38 | 1.19 | 1.43 |

The survey results indicate that the virtual breadboard served its purpose as an alternative to the physical breadboard. Furthermore, deliberate design decisions, such as the placement of the pinout diagram, helped students understand the topics being taught in the course. Moreover, while closely emulating the physical breadboard, the virtual breadboard is not hard to use, helping students become proficient in it quickly and thus saves them time on other tasks. The survey results also show the breadboard interface has room for future improvements.

## 5 Future Works

We plan to extend the UW remote laboratory to support the LTC connector and the ADC. The LTC connector on the DE1-SoC board contains the Serial Peripheral Interface (SPI) master, an $I^2C$, and a GPIO interface. The communication is enabled by the LTC connector and the hard processor system (HPS) system [7]. The real code will be passed to the ARM using the LTC connector. We are working on providing better support of ARM, for example, running bare-metal code.

We plan to support Analog-to-Digital Converter (ADC) and Digital-to-Analog Converter (DAC) on the DE1-SoC in the near future. There is two proposed usage of the ADC/DAC. Firstly, we can connect a single sensor to the FPGA, which captures the analog signal. Then we initialize the ADC to read some digital values based on the reading of the sensor. Secondly, we can connect ADC/DAC to another device, for example, the Raspberry Pi.

Combining ADC and LTC connectors, we are opening the doors for students to implement more exciting projects involving sensor readings. Additionally, we are analyzing options to improve the GPIO connectivity in the breadboard by providing virtual components, such as logic gates, to improve the power of the current solution.

## 6 Conclusion

During Autumn 2020, we taught Design of Digital Circuits and Systems at the University of Washington using the remote laboratory facility[2]. The remote laboratory facility is relying on international collaboration with other universities, including the ones described in [8]. We introduced a novel approach during this offering by using a virtual breadboard, which allows students to interact bi-directionally with the FPGA using LEDs and switches. Traditionally, FPGA remote laboratories only support using pre-wired switches and buttons, and LEDs seen on the camera. However, it does support wiring the GPIO freely. This contribution represents a first step of providing a hybrid simulated and real solution for teaching labs using GPIO in class, providing a certain degree of freedom to students compared to existing solutions. We have evaluated this approach by conducting an anonymous survey for the 60 students enrolled in the course who went through this pilot project. The results show that we were able to successfully transform a lab experiment that has been traditionally conducted in-person to an online modality. Our approach has room for improvement which our team plans to undertake in future work.

[2] https://weblab.ece.uw.edu